\documentclass[seceq]{ptptex}

\usepackage{graphicx}

\title{
Nuclear quantum transport for barrier problems%
}

\author{
Christian \textsc{Rummel} and Helmut \textsc{Hofmann}%
}

\inst{
Physik-Department der TU M\"unchen, D-85747 Garching, Germany
}

\abst{
A method is presented which allows one to introduce collective
coordinates self-consistently, in distinction to the
Caldeira-Leggett model.  It is demonstrated how the partition
function $\mathcal{Z}$ for the total nuclear system can be
calculated to deduce information both on its level density as well
as on the decay rate of unstable modes. For the evaluation of
$\mathcal{Z}$ different approximations are discussed. A recently
developed variational approach turns out superior to the
conventional methods that include quantum effects on the level of
local RPA. Dissipation is taken into account by applying energy
smearing, simulating in this way the coupling to more complicated
states. In principle, such a coupling must depend on temperature.
Previous calculations along another microscopic approach show this
fact to imply an intriguing variation of the transport
coefficients of collective motion with $T$. The relevance of this
feature is demonstrated for the thermal fission rate and for the
formation probability of super-heavy elements.}

\newcommand{\expo}{\textrm{e}}
\newcommand{\imag}{\textrm{i}}

\begin{document}

\maketitle

\section{Problems of the Caldeira-Leggett model \\
         from nuclear physics point of view}
\label{caldlegg}

The two most prominent examples of nuclear physics where processes
are governed by (iso-scalar) motion across potential barriers are
fusion and fission. In both cases dissipation may play an
important role. For large excitation energies the process is dominated
by thermal activation. With decreasing energies quantum effects
become more and more important until at zero temperature mere
quantum tunnelling is the only contribution at sub-barrier
energies.

Dissipation in the collective degrees of freedom describes flow of
energy to a large set of other, fastly relaxing degrees of
freedom, commonly referred to as the "heat bath" although in
nuclear physics this notion requires greatest care.  Sometime ago
Caldeira-Leggett have suggested a model \cite{caa.lea:ap:83a}
which allows for an exact treatment of the bath degrees of
freedom. This becomes possible at the expense of an
oversimplification of both the bath itself as well as of its
coupling to collective motion. Whereas this assumption does not
seem to be very restrictive for applications in condensed matter
physics, it violates essential requirements for nuclear collective
motion \cite{hoh:pr:97}. Let us just mention the most serious
deficiencies.

\begin{itemize}
\item In the nuclear case the collective degrees of freedom (CDOF)
  are {\em not} independent of the nucleonic ones, viz of the heat bath.
  Already the simplest
  condition on self-consistency implies that the shape of the mean field
  in which the nucleons move must vary in non-linear way with the CDOF.
\item Even in the simple case of small amplitude collective motion,
  the frequencies are determined by RPA-like secular equations
  which differ essentially from those of the Caldeira-Leggett model
  \cite{hoh:pr:97}.
\item Related to this issue is the fact that the Caldeira-Leggett
  model is unable to make predictions for the evaluation of the transport
  coefficients for collective motion like potential energy, inertia and
  even for the strength of friction. All of them depend sensitively on both
  the  collective coordinate as well as on temperature.
\end{itemize}

Without invoking the Caldeira-Leggett model so far quantum effects
in nuclear dissipative dynamics were mainly treated applying real
time propagation \cite{hoh:pr:97,hoh.ivf:prl:99}, with the
exception of \cite{ruc.hoh:pre:01}.
Whereas the former approach is based on the deformed shell model
in the novel approach one starts with a typical nuclear
Hamiltonian of two body nature {\em before} CDOF are introduced,
as shall be explained below.

\section{Self-consistent dynamics for separable two body Hamiltonians}
\label{septwobodham}

Suppose we are given the Hamiltonian with separable two body
interaction of
\begin{equation} \label{twobodham}
\hat{\cal H} = \hat{H} + \frac{k}{2} \,\hat{F}\hat{F} \qquad
\textrm{with} \qquad k < 0\,.
\end{equation}
It may be understood as one term in an expansion of the two body
interaction into separable terms. For transport problems this
ansatz has been used before applying the Bohm-Pines method of
introducing collective coordinates, see \cite{hoh:pr:97} with
further references given therein. These applications were based on
a real time approach, implying that this method is limited to
temperatures above a certain $T_{c}$. Here, we want to follow an
imaginary time approach
\cite{pug.bop.brr:ap:91,ath.aly:npa:97,ruc.hoh:pre:01}. Different
to the Bohm-Pines method the collective variable is introduced
through a Hubbard-Stratonovich transformation in the functional
integral for the partition function of (\ref{twobodham}) we arrive
at the one body Hamiltonian
\begin{equation} \label{onebodham}
\hat{\cal H}_\textrm{HS} =
\hat{H} + q(\tau) \hat{F} + \frac{1}{2|k|} q^{2}(\tau)
\end{equation}
where the collective coordinate $q$ depends on the imaginary time
$\tau = 0 \ldots \hbar/T$ and has been introduced self-consistently.
After expanding the periodic collective ``path'' into a Fourier series
$q(\tau) = q_{0} + \sum_{r \ne 0} q_{r} \,\exp (\imag\nu_{r}\tau)$
where $\nu_{r} = (2\pi/\hbar) Tr$ are the Matsubara frequencies
the final form of the functional integral for the partition function
reads
\begin{equation} \label{partfunc}
{\cal Z}(T) = \frac{1}{\sqrt{2\pi |k|T}}
\int dq_{0} \ \exp (-{\cal F}^\textrm{SPA}(T, q_{0}) / T)
\,C(T, q_{0}) \,.
\end{equation}
The $q_{0}$-integral represents the {\em thermal} fluctuations, while
the {\em quantum} fluctuations with amplitudes $q_{r}$ that become
important at low temperatures enter the remaining path integral for
the ``dynamical'' corrections
\begin{equation} \label{corrgen}
C(T, q_{0}) =
\int {\cal D}'q \ \exp \left( -\frac{A(T, q_{0})}{|k|T} \right)
\end{equation}
where the ``action'' can be expanded like
\begin{equation} \label{expandA}
A(T, q_{0}) =
  \lambda_{rs}(T, q_{0}) q_{r}q_{s} + \rho_{rst}(T, q_{0}) q_{r}q_{s}q_{t}
+ \sigma_{rstu}(T, q_{0}) q_{r}q_{s}q_{t}q_{u} + {\cal O}(q_{r}^{5}) \,.
\end{equation}
The exponential in (\ref{partfunc}) represents the partition
function of the static part of the Hamiltonian (\ref{onebodham})
with ${\cal F}^\textrm{SPA}(q_{0})$ being the corresponding free
energy. The simplest approximation to the dynamical corrections
consists in neglecting them entirely: $C^\textrm{SPA}(T, q_{0})
\equiv 1$ represents the classical limit and has been given the
name Static Path Approximation (SPA) in the past 
\cite{mub.scd.der:prb:72}.

The coefficient $\lambda_{rs}$ of the second order part of
(\ref{expandA}) can be expressed in terms of the response function
$\chi(T, q_{0}, \omega)$ that describes the changes of the average
$\langle\hat{F}\rangle_\omega$ for small variations of the path
$q(\omega)$ \cite{ruc.hoh:pre:01}:
\begin{equation} \label{lambdachi}
\lambda_{rs}(T, q_{0}) =
\left( 1 + k \chi(T, q_{0}, \omega) \right) \,\delta_{r,-s} =
\frac{\prod_{\mu} (\nu_{r}^{2} + \omega_{\mu}^{2}(T, q_{0}))}
     {\prod_{k>l} (\nu_{r}^{2} + \omega_{kl}^{2}(q_{0}))} \,\delta_{r,-s}
\end{equation}
Here, the $\omega_{\mu}$ are the local RPA frequencies and
$\hbar\omega_{kl} = \epsilon_{k} - \epsilon_{l}$ are the energies
of the intrinsic excitations associated to the static part of the Hamiltonian
(\ref{onebodham}). In cases where $\lambda_{rs}$ is positive and
sufficiently large a truncation of (\ref{expandA}) after the
second order will do and the remaining $q_{r \ne 0}$-integrals
in ${\cal D}'q$ can be evaluated exactly. One ends up with the
result of the Perturbed Static Path Approximation (PSPA)
\begin{equation} \label{C-PSPA}
C^\textrm{PSPA}(T, q_{0}) = \prod_{r>0} \lambda_{r}^{-1}(T, q_{0})
\end{equation}
which takes into account quantum fluctuations on the level of
local RPA \cite{pug.bop.brr:ap:91,ath.aly:npa:97,ruc.hoh:pre:01}.
As $\lambda_{rs}$ depends on temperature the condition 
$\lambda_{1} > 0$ for the breakdown of this approximation due to diverging
integrals can be transformed into the requirement $T > T_{0}$, where for the
``crossover temperature'' (note the similarity to the case of
dissipative tunneling \cite{disstunn})
$T_{0} < T_{c}$ holds true \cite{ruc.hoh:pre:01}.

The coefficients $\rho_{rst}$ and $\sigma_{rstu}$ of (\ref{expandA})
have been calculated in \cite{ruc.anj:epjb:02}
where the breakdown of the approximation has
been shifted to $T_{0}/2$ by treating the $q_{1}$- and
$q_{2}$-integrals in ${\cal D}'q$ to fourth order analytically.
Meanwhile an approximation has been developed
\cite{ruc.hoh:var} that takes over basic ideas of a
variational principle suggested for one-dimensional systems
in \cite{giachetti} and \cite{fer.klh:pra:86}.
Choosing a suitable $A_\Omega^{q_{0}}$ the correction factor
(\ref{corrgen}) can be rewritten as the average defined with respect
to the weight $\exp ( -A_\Omega^{q_{0}}/|k|T)$
\begin{equation} \label{corraverage}
C(T, q_{0}) =
\left\langle \exp \left( -\frac{A(T, q_{0}) - A_\Omega^{q_{0}}}{|k|T}
\right) \right\rangle \,.
\end{equation}
In order to determine the best approximation to the equilibrium
quantity ${\cal Z}(T)$ the convexity of the exponential
$\langle \expo^{-x} \rangle \ge \expo^{-\langle x \rangle}$
can be used such that the following inequality can be exploited:
\begin{equation} \label{corrvar}
C(T, q_{0}) \ge C^\textrm{var}(T, q_{0}) =
\exp \left( -\frac{\langle A(T, q_{0}) - A_\Omega^{q_{0}}
\rangle}{|k|T} \right)
\end{equation}
For $A_\Omega^{q_{0}}$ the PSPA form is chosen but with the RPA
frequencies $\omega_{\mu}(T, q_{0})$ replaced by a number of
{\em adjustable parameters} $\Omega_{\mu}(T, q_{0})$. Maximizing the
right hand side of (\ref{corrvar}) with respect to
$\Omega_{\mu}(T, q_{0})$ gives an optimal approximation to the
partition function (\ref{partfunc}).
\begin{figure}
\centerline{\includegraphics[width=60mm,angle=-90]{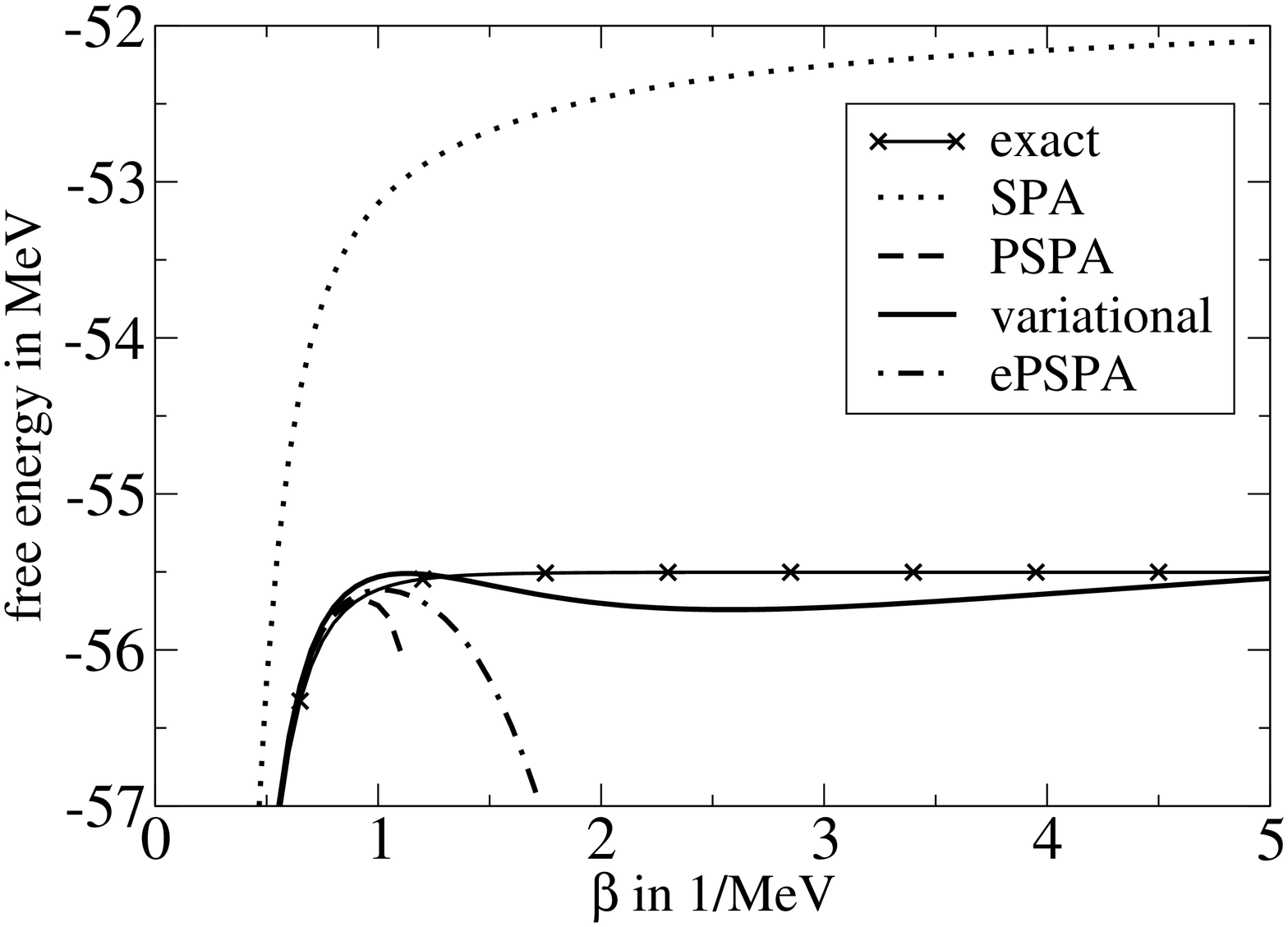}
\includegraphics[width=60mm,angle=-90]{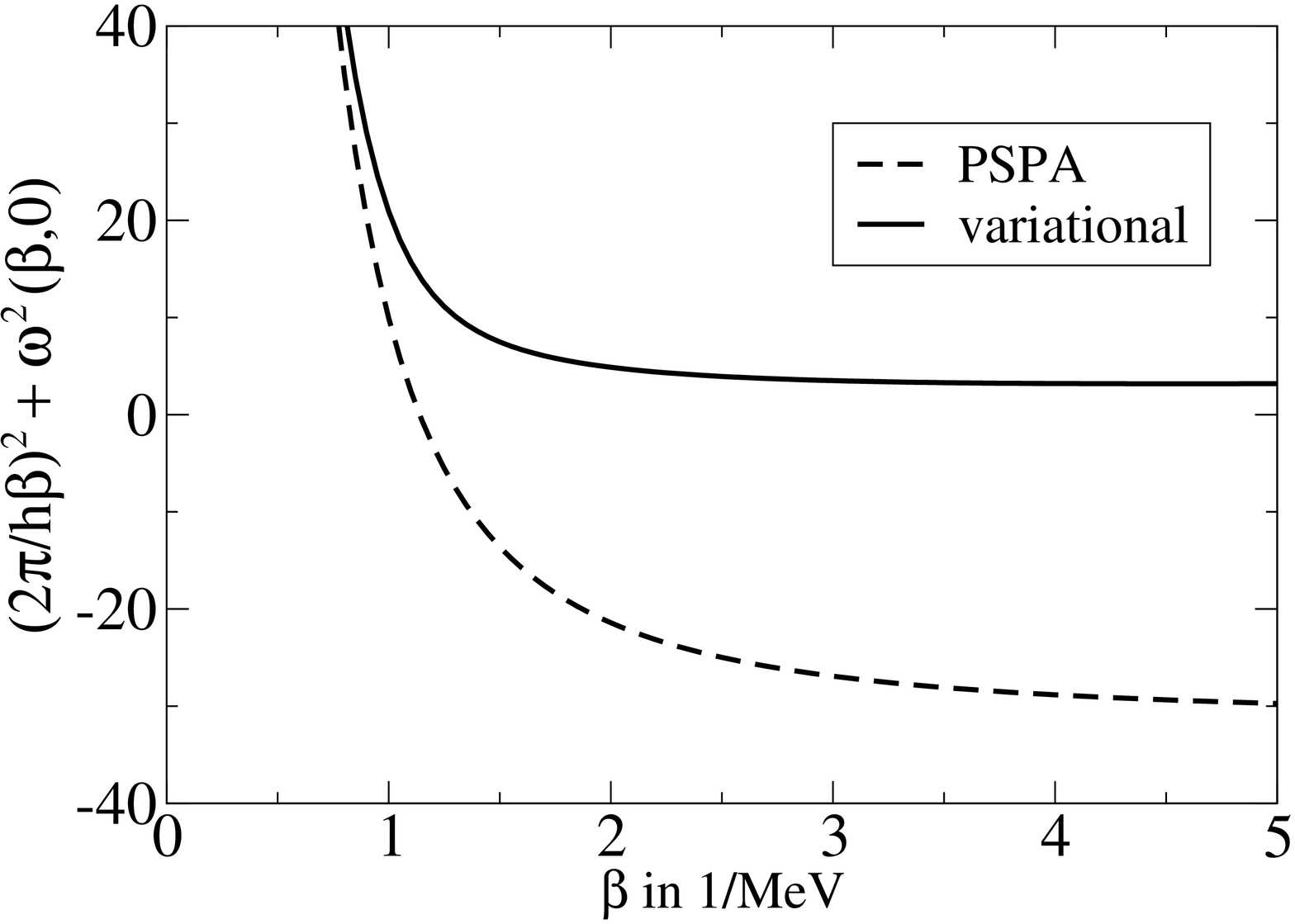}}
\caption{\label{fig-freeen}
Left: Free energy of the LMG in various approximations.
The variational principle gives very good results up to
$\beta = 1/T = 5 \,\textrm{MeV}^{-1}$ whereas the PSPA brakes down at
$\beta_{0} = 1/T_{0} = 1.13 \,\textrm{MeV}^{-1}$.
Right: $\beta$-dependence of the quantities
$\nu_{1}^{2} + \omega_\textrm{RPA}^{2}(q_{0}=0)$ and
$\nu_{1}^{2} + \Omega_\textrm{var}^{2}(q_{0}=0)$.}
\end{figure}
In the left panel of fig.~\ref{fig-freeen} we demonstrate at the
example of the free energy of the exactly solvable
Lipkin-Meshkov-Glick model (LMG) that the variational approach
gives excellent results and is applicable even at {\em very} low
temperatures. The classical SPA is good only at high temperatures.
Inclusion of quantum effects on the RPA level via PSPA improves
the approximation at high temperatures considerably but breaks
down at $T_{0}$. The approximation of \cite{ruc.anj:epjb:02}, 
called ePSPA, behaves well in the crossover region 
$T \approx T_{0}$. In order
to illustrate the reason for the applicability of the variational
principle at low temperatures we plot in the right panel of
fig.~\ref{fig-freeen} the $\beta$-dependence of the quantities
$\nu_{1}^{2} + \omega_\textrm{RPA}^{2}(T, q_{0}=0)$ and
$\nu_{1}^{2} + \Omega_\textrm{var}^{2}(T, q_{0}=0)$ that determine
the sign of $\lambda_{1}$ and its variational analog via 
(\ref{lambdachi}). The former
quantity becomes negative at $\beta_{0} = 1/T_{0}$ implying a
breakdown of the PSPA, whereas the second one stays positive.

\section{Introduction of dissipation and
         microscopic transport coefficients}
\label{dissipation}

So far dissipation has not been taken into account. The
coefficient $\lambda_{r}(T, q_{0})$ of (\ref{lambdachi}) contains
the response function $\chi(\omega) = \chi'(\omega) + \imag
\chi''(\omega)$. In the independent particle model the dissipative
part typically has the form (for $\omega >0$)
\begin{equation} \label{chiipm}
\chi''(\omega) = \sum_{k>l} f_{kl} \,\delta(\omega - \omega_{kl})
\end{equation}
with strengths $f_{kl}$. We know that (at least at not too small
excitations) the simple states of the independent particle model
are coupled to more complicated ones via the residual
interactions. The spectrum becomes more dense and the self-energy
of the individual states acquires a finite width such that one
ends up with a continuous function $\chi''(\omega)$. We do not go
into these details \cite{hoh:pr:97} here but for simplicity mimic
such effects by smearing out the spectrum by hand
\cite{ruc.hoh:pre:01}. To such an averaged spectrum we fit the
Lorentzian
\begin{equation} \label{Lorentzian}
\chi''(T, q_{0}, \omega) =
F^{2}(T, q_{0}) \left( \frac{\Gamma(T, q_{0})/2}
{(\omega - {\cal E}(T, q_{0}))^{2} + (\Gamma(T, q_{0})/2)^{2}}
- ({\cal E} \leftrightarrow -{\cal E}) \right)
\end{equation}
and extract the three parameters $F^{2}$, ${\cal E}$ and $\Gamma$.
In this way the full response function may be interpreted locally
as the one of a damped harmonic oscillator:
\begin{equation} \label{chiosc}
\chi(T, q_{0}, \textrm{Im} z \ne 0) = \chi_\textrm{osc} =
\frac{1}{M(T, q_{0})} \,\frac{1}{\omega_\textrm{nucl}^{2}(T, q_{0})
- \imag \Gamma(T, q_{0}) z - z^{2}}
\end{equation}
The transport coefficients for the local inertia $M$, the local
nucleonic frequency $\omega_\textrm{nucl}$ and the local damping
$\Gamma$ can easily be related to the fit parameters
\cite{ruc.hoh:pre:01} and {\em depend on temperature $T$ and the
collective coordinate $q_{0}$} simply because the original response
function has those dependencies. One can easily convince oneself that
the solution of the secular equation $1 + k \chi(\omega) = 0$ for the
collective frequencies now has a finite imaginary part, indicating that
collective motion is damped. The big advantage of the method
suggested here is that all three transport coefficients $M$,
$\omega_\textrm{nucl}$ and $\Gamma$ are {\em taken from the same
microscopic theory}, instead of piecing together various macroscopic
pictures. 

\begin{figure}
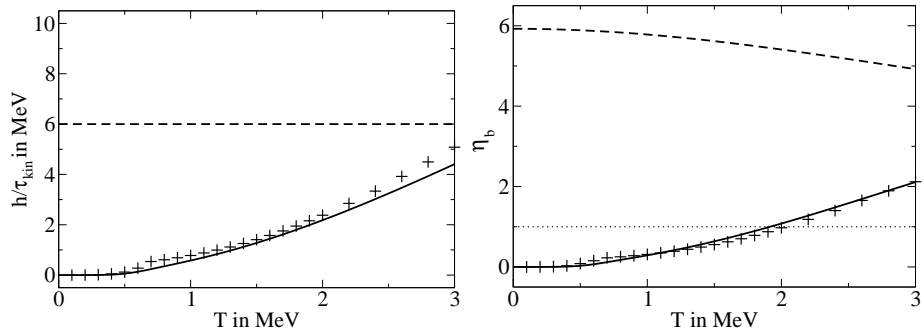

\centerline{\includegraphics[width=60mm,angle=0]{Gammab_kin_pair_publ.eps}
\includegraphics[width=60mm,angle=0]{etab_pair_publ.eps}}
\caption{\label{fig-macromicro}
Comparison of macroscopic (dashed) and microscopic (symbols) ratios of
transport coefficients for the barrier of $^{224}\textrm{Th}$.
The fully drawn line shows the simple
analytical formulas proposed in \cite{hoh.ivf.ruc.yas:prc:01}.
Left: $\Gamma = \gamma/M = \hbar/\tau_\textrm{kin}$, often called
$\beta$ in the literature.
Right: The effective damping $\eta$. The dotted line indicates the
change from under-damped ($\eta < 1$) to over-damped ($\eta > 1$)
motion.}
\end{figure}

In order to illustrate the drastic differences between
our microscopic approach and the frequently used combination of wall
friction, liquid drop stiffness and irrotational flow inertia we plot
in fig.~\ref{fig-macromicro} the temperature dependence of some ratios
of the transport coefficients for the barrier of $^{224}\textrm{Th}$.
In the macroscopic picture the ratio $\Gamma = \gamma/M$ does not
depend on temperature and is much larger than in the microscopic
picture where it increases with temperature and vanishes at $T =
0$. Within this macroscopic picture the effective damping strength
$\eta = \gamma / (2 \sqrt{|C| M})$ is strongly over-damped
implying the applicability of the Smoluchowsky limit. For the
microscopic transport coefficients, on the other hand, we observe a
transition from under-damped motion ($\eta < 1$) to over-damped
motion ($\eta > 1$); note that in the former case the inertia
plays an important role. Even at $T = 3 \,\textrm{MeV}$ the
Smoluchowsky limit is not fully reached.

\section{Application I: Thermal fission}
\label{fission}

Applying a saddle point approximation to the $q_{0}$-integral in
(\ref{partfunc}) we can calculate the rate of thermal fission from the
imaginary part of the free energy from the formula
\cite{laj:ap:67,afi:prl:81}
\begin{equation} \label{ImF}
R(T) = -\frac{2}{\hbar} \,\frac{T_{0}}{T} \,\textrm{Im} \mathcal{F}(T)
\qquad \textrm{for} \qquad T > T_{0} \,.
\end{equation}
In situations where the  potential is given by just one minimum
located at $q = q_{a}$ and just one parabolic barrier at $q =
q_{b}$ we obtain
\begin{equation} \label{rate}
R = \frac{\varpi(q_{b})}{2\pi}
\ \sqrt{\frac{\mathcal{C}_\mathcal{F}(q_{a})}
{|\mathcal{C}_\mathcal{F}(q_{b})|}}
\ \left( \sqrt{1 + \eta^{2}(q_{b})} - \eta(q_{b}) \right)
\ \expo^{-\mathcal{F}^{\textrm{SPA}}(q_{b}) / T}
\cdot \frac{C(q_{b})}{C(q_{a})} \,.
\end{equation}
In order to simplify notation we have omitted the $T$-dependence
everywhere. Besides the Arrhenius factor there appear non-trivial
ones which are worth to be discussed in detail. The first two
factors can be rewritten in the form $\varpi(q_{a})/2\pi
\cdot \sqrt{M(q_{a})/M(q_{b})}$ where the attempt frequency at the
minimum and the ratio of the inertias at the minimum and the
barrier appear. The term in brackets is the famous Kramers factor
\cite{krh:ph:40} that takes into account the decrease of the decay
rate due to damping. The last factor represents the increase of
the rate due to quantum corrections.  It has been calculated in
\cite{frp.tig:npa:92} and \cite{hoh.ivf:prl:99} 
for the example of $^{224}\textrm{Th}$. While \cite{frp.tig:npa:92}
use a Caldeira-Leggett type model, 
microscopic temperature- and coordinate dependent transport
coefficients are used in \cite{hoh.ivf:prl:99}. 
Its contribution turns out to be significant at low
temperatures (but with $T > T_{0}$).

\begin{figure}
\centerline{\includegraphics[width=60mm,angle=-90]{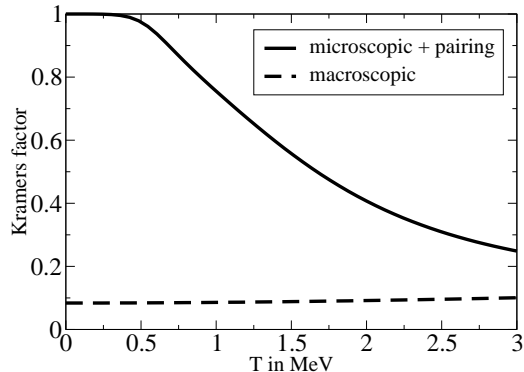}}
\caption{\label{fig-Kramers}
$T$-dependence of Kramers' factor for macroscopic and microscopic
  transport coefficients.}
\end{figure}
In \cite{hoh.ivf.ruc.yas:prc:01} the importance of microscopic
transport coefficients for the description of thermal fission has
been examined in more detail. It has been found that the
experimentally observed phenomenon of ``onset of dissipation''
\cite{thm.beg:prl:93} cannot be explained by the macroscopic set
of transport coefficients. In order to illustrate this statement
we plot in fig.~\ref{fig-Kramers} the temperature dependence of
Kramers' factor. Using macroscopic transport coefficients no
``onset of dissipation'' can be seen at all, whereas for
microscopic ones it appears around the $T \approx 0.5
\,\textrm{MeV}$, exactly where pairing disappears.

\section{Application II: Formation probabilities in fusion}
\label{formation}

Now we like to turn to a more recent application of microscopic 
$T$- and $q_{0}$-dependent transport coefficients to the
probability of compound nucleus formation in production of
super-heavy elements \cite{ruc.hoh:npa:03}. We do not study the
initial phase of the reaction explicitly but make the assumption
that due to dissipation the sum of the incident energy $E$ and the
assumed barrier height $E_{b} = 10 \,\textrm{MeV}$ is distributed
into a remaining kinetic energy and a thermal excitation according
to
\begin{equation} \label{splitenergy}
E_{0}^\textrm{kin} = R (E + E_{b})
\qquad \textrm{and} \qquad
E_{0}^{*} = (1 - R) (E + E_{b}) \,.
\end{equation}
For the parameter $R$ we compare the effect of the two assumptions
$R = 0$ and $R = 0.25$ on the final formation probability
\begin{equation} \label{formprob}
\Pi_\infty = \frac{1}{2} \ \textrm{erfc}
\left( -\frac{q_\infty}{\sqrt{2 \Sigma_{qq}^\infty}} \right)
\end{equation}
with
\begin{equation} \label{qinf}
q_\infty = \sqrt{\frac{E_{0}^\textrm{kin}}{\hbar\varpi_{b}}} +
\sqrt{\frac{E_{b}}{\hbar\varpi_{b}}} \,\frac{z^{-}}{\varpi_{b}}
\qquad \textrm{where} \qquad
\frac{z^{-}}{\varpi_{b}} = -\frac{\sqrt{1 + \eta_{b}^{2}} + \eta_{b}}{2}
\end{equation}
and
\begin{equation} \label{sigmainf}
2 \Sigma_{qq}^\infty =
\frac{\Sigma_{pp}(t_{0}) - \Sigma_{pp}^{eq,b}}{\hbar^{2}} +
\left( \frac{z^{-}}{\varpi_{b}} \right)^{2}
\left( \Sigma_{qq}(t_{0}) - \Sigma_{qq}^{eq,b} \right) \,.
\end{equation}
\begin{figure}
\centerline{\includegraphics[width=60mm,angle=-90]{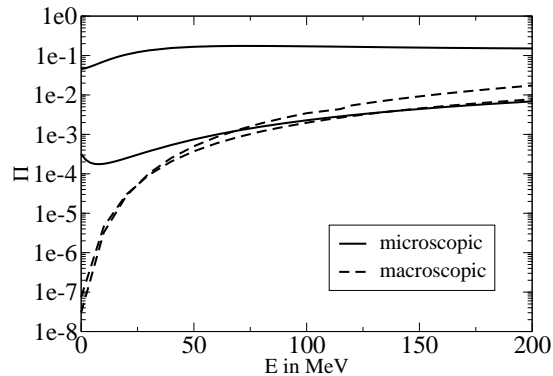}}
\caption{\label{fig-formprob}
Formation probability as a function of incident energy.
The initial conditions $\Sigma_{pp}(t_{0}) = \Sigma_{pp}^{eq,b}$
of thermal equilibrium for the momentum distribution,
$\Sigma_{qq}(t_{0}) \Sigma_{pp}(t_{0}) = 3\hbar^{2}/4$ and
$\Sigma_{qp}(t_{0}) = 0$ are used.
Clearly these fluctuations are compatible with quantum mechanics.}
\end{figure}
In fig.~\ref{fig-formprob} we compare the dependence of the formation
probability on incident energy for microscopic
transport coefficients with that of macroscopic ones. In both cases
the upper curve corresponds to finite initial kinetic energy
($R = 0.25$) whereas the lower one represents $R = 0$.
Most striking is the fact that for not too high incident energies
the weak damping in the case of
microscopic transport coefficients results in formation probabilities
that are {\em orders of magnitude larger} than those of the
macroscopic picture. In addition to that -- as expected for the case of
under-damped or slightly over-damped motion
$\eta_{b} \approx 0.5 \ldots 2$ -- the microscopic formation
probability is extremely sensitive to the initial kinetic energy. Such a
dependence is very weak in the case of macroscopic transport
coefficients. The reason is found in the fact that in this case motion
is strongly over-damped $\eta_{b} \approx 5 \ldots 6$.

\section{Summary}
\label{summary}

We have developed a method for the description of dissipative
collective motion in finite nuclei that does not suffer from the
deficiencies of the Caldeira-Leggett model. This method enables
one to calculate microscopically on the same footing transport
coefficients for inertia, damping and stiffness, including their
dependence on temperature and collective coordinate. Comparison
with the macroscopic picture of the liquid drop model and wall
friction shows significant differences, resulting in much weaker
damping at low temperatures in general. These effects are shown to
be important in thermal fission as well as in the formation of the
compound nucleus in super-heavy element production.

\section*{Acknowledgments}
The authors would like to thank F.A.~Ivanyuk for critical and fruitful
discussions.

%



\end{document}